\documentclass[a4paper,12pt,russian]{article}
\usepackage{cmap}
\usepackage[T2A]{fontenc}
\usepackage[utf8]{inputenc}
\usepackage[russian,english]{babel}
\usepackage{amssymb}
\usepackage{ifthen}
\usepackage{color}
\usepackage{hyperref}

\makeatletter
\def\@seccntformat#1{\csname the#1\endcsname.\ } 
\def\@biblabel#1{[#1]} 
\makeatother

\date{}

\newenvironment{proof}[1][\hspace{-1.0ex}]%
{\par\addvspace{1mm}\emph{Proof\hspace{1.0ex}{#1}.} }%
{\quad$\blacktriangle$\par\addvspace{1mm}}

\newif\ifNoRemark
\def\addtheorem#1#2#3#4{
\ifthenelse{\equal{#2}{}}{}%
{\ifthenelse{\expandafter\isundefined\csname the#2\endcsname}{\newcounter{#2}}{}}
\newenvironment{#1}[1][\global\NoRemarktrue]% No Remark by default
{\par\addvspace{2mm plus 0.5mm minus 0.2mm}\noindent % new paragraph without indent
{\bf #3}\ifthenelse{\equal{#2}{}}{}%
{\refstepcounter{#2}{\bf ~\csname the#2\endcsname}}%
{\bf \vphantom{##1}\ifNoRemark.\ \else\ (##1).\fi}\begingroup #4}%
{\endgroup\par\addvspace{1mm plus 0.5mm minus 0.2mm}\global\NoRemarkfalse}
\expandafter\newcommand\csname b#1\endcsname{\begin{#1}}
\expandafter\newcommand\csname e#1\endcsname{\end{#1}}
}

\addtheorem{theorem}{theorem}{Theorem}{\sl}
\addtheorem{proposition}{prpst}{Proposition}{\sl}
\addtheorem{lemma}{lemma}{Lemma}{\sl}
\addtheorem{corollary}{corollary}{Corollary}{\sl}
\addtheorem{remark}{}{Remark}{\rm}

\begin{document}

\title{Constructions of transitive latin hypercubes% 
%\thanks{The work was funded by the Russian Science Foundation (grant No 14-11-00555).}
}
\author{Denis S. Krotov and Vladimir N. Potapov}

\maketitle

\begin{center}
\textit{Sobolev Institute of Mathematics, 
pr. Akademika Koptyuga 4, 
Novosibirsk 630090, Russia}
\end{center}

\begin{abstract}
A function $f:\{0,...,q-1\}^n\to\{0,...,q-1\}$ invertible in each argument is called a latin hypercube.
A collection $(\pi_0,\pi_1,...,\pi_n)$ of permutations of $\{0,...,q-1\}$ is called an autotopism of a latin hypercube $f$
if $\pi_0f(x_1,...,x_n)=f(\pi_1x_1,...,\pi_nx_n)$ for all $x_1$, ..., $x_n$.
We call a latin hypercube isotopically transitive (topolinear) if its group of autotopisms acts transitively (regularly)
on all $q^n$ collections of argument values.
We prove that the number of nonequivalent topolinear latin hypercubes grows exponentially with respect to $\sqrt{n}$
if $q$ is even and exponentially with respect to $n^2$ if $q$ is divisible by a square.
We show a connection of the class of isotopically transitive latin squares with the class of G-loops, known in noncommutative algebra,
and establish the existence of a topolinear latin square that is not a group isotope.
We characterize the class of isotopically transitive latin hypercubes of orders $q=4$ and $q=5$.

\end{abstract}

We consider the latin hypercubes such that 
their autotopism groups
act transitively (or regularly) on their elements.
We call them the isotopically transitive (topolinear, respectively) latin hypercubes.
The study of highly symmetrical objects, such as the objects from the considered class,
is a natural direction in the enumerative combinatorics.
On the other hand, latin hypercubes are also very natural research objects, which are
studied in different areas of mathematics. For example, in coding theory, equivalent
objects are known as the distance-$2$ MDS codes; in noncommutative algebra, the $n$-ary quasigroups.
The number of latin $n$-cubes, MDS codes, and related objects attracts attention of mathematicians
in the last few years; different evaluations and exact values can be found in
\cite{DonGra:2013},
\cite{Ito:patent},
\cite{KKO:smallMDS},
\cite{KokOst:Gr-Lat},
\cite{KokOst:further},
\cite{KPS:ir},
\cite{LinLur:2014},
\cite{MK-W:small}, 
\cite{PotKro:numberQua.en},
\cite{Pot:Number-MDS}.
In this paper, we are mainly concentrated on the number of nonequivalent
isotopically transitive latin hypercubes,
proving that this number is exponential with respect to the dimension,
for some fixed orders.
However, some additional interesting characterization results are obtained for isotopically transitive  latin squares
and isotopically transitive latin hypercubes of order $4$.

In Section~\ref{s:P}, we give definitions and some preliminary results.
In Section~\ref{s:itG}, Theorem~\ref{th:IT=G}, we prove a connection (in some sense, one-to-one correspondence)
 between the isotopically transitive latin squares and the algebraic structures known as G-loops.
In Section~\ref{s:G}, we consider examples of G-loops,
which are utilized in Section~\ref{s:I} to construct an exponential (in $\sqrt n$)
number of nonequivalent topolinear latin $n$-cubes,
for every even order $\ge 4$ (Theorems~\ref{th:iter1} and~\ref{th:iter2}, Corollary~\ref{cor:No}).
Additionally, in Section~\ref{s:G} we show that there are topolinear latin squares
that are not isotopic to a group operation (Corollary~\ref{c:C2p}).
In Section~\ref{s:Q}, we consider a direct construction, using quadratic functions, which gives
an exponential (in $n^2$) number of nonequivalent topolinear latin $n$-cubes,
for every order divisible by a square (Corollary~\ref{cor:No2}).
In Sections~{\ref{s:4}} and~{\ref{s:5}}, we characterize the class of isotopically transitive latin hypercubes of orders $4$ and $5$ 
(Theorem~\ref{th:order4} and~\ref{th:5}, respectively).
In the concluding section, we consider some open problems.

%=====================================================================
\section{Preliminaries}\label{s:P}
\subsection{Transitive and propelinear sets}\label{ss:trans}

Let $\Sigma=\Sigma_q$ be a finite set of cardinality $q$;
for convenience, we choose some element of $\Sigma$ and denote it $0$.
The set $\Sigma^n$ of $n$-tuples from $\Sigma^n$
with the Hamming distance is called a $q$-\emph{ary} $n$-{\it
dimensional Hamming space}
(recall that the Hamming distance between two $n$-tuples is the number of positions in which they differ).
An \emph{isotopism} $\overline{\tau}= (\tau_0,\ldots,\tau_{n-1})$ is a transform
$\overline{x}\mapsto\overline{\tau}\overline{x}$, where
$\overline{x}=(x_0,\ldots,x_{n-1})\in \Sigma^n$,
$\overline{\tau}\overline{x}=(\tau_0x_0,\ldots,\tau_{n-1}x_{n-1})$,
and
$\tau_0$, \ldots, $\tau_{n-1}$ are permutations of $\Sigma$.
For a set $A\subseteq \Sigma^n$,
denote
$\overline{\tau}A= \{\overline{\tau}\overline{x}\ | \ \overline{x}\in A\}.$
Define the  \emph{autotopism group}
$\mathrm{Ist}(A)=\{\overline{\tau} \ |\ \overline{\tau}A=A\}$,
which consists of isotopisms that  map $A\subseteq \Sigma^n$ to itself.
It is well known (see, e.g., \cite[Theorem 9.2.1]{Brouwer}) that every isometry of $\Sigma^n$
can be represented as the composition of an isotopism and a coordinate permutation.
The subgroup of the isometry group of $\Sigma^n$ that maps
$A\subseteq \Sigma^n$ to itself will be denoted $\mathrm{Aut}(A)$.
We will say that two subsets $A$ and $B$ of $\Sigma^n$ are \emph{equivalent} (\emph{isotopic})
if $B$ is the image of $A$ under some isometry of the space (isotopism, respectively).

A set $A\subseteq \Sigma^n$ is called
  \emph{transitive} if for every two vertices
  $\overline{x}$, $\overline{y}$ from $A$ there exists
   an element $\alpha$ of $\mathrm{Aut}(A)$
such that
 $\alpha(\overline{x})=\overline{y}$; i.e.,
 the group $\mathrm{Aut}(A)$ acts transitively on $A$.
We call a set $A\subseteq \Sigma^n$
 \emph{isotopically transitive} if $\mathrm{Ist}(A)$ acts transitively on $A$.
In what follows we assume that the all-zero tuple $\overline{0}$ belongs to $A$.
Note that to make sure that $A\subseteq \Sigma^n$ is (isotopically) transitive,
it is sufficient to check that the condition of the definition holds for some $\overline{x}\in A$,
say $\overline{x}=\overline{0}$, and all $\overline{y}$, or for some $\overline{y}$ and all $\overline{x}$.

\begin{remark}
 For $|\Sigma|=2$,
the isotopically transitive sets are exactly the affine subspaces of $\Sigma^n$,
considered as a vector space over the field $\mathrm{GF}(2)$.
\end{remark}

 A set $A\subseteq \Sigma^n$ is called \emph{propelinear} \cite{RBH:89} if
 $\mathrm{Aut}(A)$ includes a regular subgroup, i.e., a subgroup of
 $\mathrm{Aut}(A)$ of cardinality $|A|$ that acts transitively on $A$.
  We call a set $A\subseteq \Sigma^n$ \emph{topolinear} if
 $\mathrm{Ist}(A)$ includes a regular subgroup $G_A$.

Let $A$ be a subset of $\Sigma^n$. By a \emph{subcode} $R$ of $A$,
we will mean a subset of $\Sigma^{m}$, $m\le n$, obtained from $A$
by ``fixing'' $n-m$ coordinates.
We explain this by defining a subcode recursively.
Define an $(n-1)$-subcode of $A$ as the set
$\{(x_0,\ldots,x_{j-1},x_{j+1},\ldots,x_{n-1})\mid (x_0,\ldots,x_{j-1},a,x_{j+1},\ldots,x_{n-1})\in A\}$
for some $j\in \{0,\ldots,{n-1}\}$ and $a\in \Sigma$;
then, for $m<n$, an $m$-subcode (or simply a subcode) of $A$
is defined as $((m{+}1)-1)$-subcode of an $(m+1)$-subcode of $A$ (the set $A$ itself is an $n$-subcode).

\begin{proposition}\label{p:subcodes-of-IT}
{\rm 1)} The subcodes of an isotopically transitive set are isotopically transitive.

{\rm 2)} The subcodes of a topolinear set are topolinear.
\end{proposition}

\begin{proof}
1) Let $A\subseteq \Sigma^n$ be an isotopically transitive set, and let $R$ be an $m$-subcode of $A$.
Without loss of generality we can assume that
$$R=\{(x_0,\dots,x_{m-1}) \ |\ (x_0,\dots,x_{m-1},0,\ldots,0)\in A\} \ni (0,\ldots,0).$$
For given $(x_0,\ldots,x_{m-1})\in R$,
there is $\overline\tau = (\tau_0,\ldots,\tau_{n-1})\in \mathrm{Ist}(A)$ such that
$\overline\tau \overline{0}=(x_0,\ldots,x_{m-1},0,\ldots,0)$.
Then
$\tau_{m}(0)=\dots=\tau_{n-1}(0)=0$;
hence, $\overline\tau'= (\tau_0,\ldots,\tau_{m-1})$
is an autotopism of $R$ that sends $(0,\ldots,0)$ to $(x_0,\ldots,x_{m-1})$.

2) If $\mathrm{Ist}(A)$ has a subgroup $G$ that acts regularly on $A$, then for every $(x_0,\ldots,x_{m-1})\in R$
the choice of $\overline\tau\in G$ is unique,
and the corresponding autotopisms $\overline\tau'$ form a group of cardinality $|R|$ that acts transitively on $R$.
\end{proof}

For two sets $A\subseteq \Sigma^n$ and $B\subseteq \Theta^n$, 
their Cartesian product $A\times B \subseteq (\Sigma\times\Theta)^n$ is defined as
$A\times B = \{([a_0,b_0],\ldots,[a_{n-1},b_{n-1}]) \ |\ (a_0,\ldots,a_{n-1})\in A, (b_0,\ldots,b_{n-1})\in B \}$.
The following statement is straightforward.

\begin{proposition}\label{p:cartesian-product}
Let $A\subseteq \Sigma^n$ and $B\subseteq \Theta^n$ are
isotopically transitive (topolinear) sets.
Then the set $A\times B \subseteq (\Sigma\times\Theta)^n$ is isotopically transitive (topolinear).
\end{proposition}

\begin{remark}
For the transitive sets in general,
the statements similar to Propositions~\ref{p:subcodes-of-IT} and~\ref{p:cartesian-product}
do not hold.
For example,
the set
$\{001,011,010,110,100,101\}\subset \Sigma^3=\{0,1\}^3$
is propelinear,
while all its $2$-subcodes
(e.g., $\{01,11,10\}$)
are not transitive.
The sets $\{001,010,100\}$ and $\{000,001\}$ are propelinear,
while their Cartesian product is not transitive.
\end{remark}

%-------------------------------------------------------------------
\subsection{MDS codes, latin hypercubes}\label{ss:MDS}
A set $M\subset \Sigma^n$ is called an \emph{MDS code}
(with code distance $2$) of length $n$ if $|M|=q^{n-1}$
and the distance between any two different code tuples is at least $2$.
A function $f:\Sigma^n\rightarrow \Sigma$ is called
a \emph{latin $n$-cube} (a \emph{latin hypercube}; in the case $n=2$, a \emph{latin square}) of order $q$
if $f(\overline{x})\neq f(\overline{y})$ holds for every two neighbor ($d(\overline{x},\overline{y})=1$) vertices
$\overline{x},\overline{y}\in \Sigma^n$.
(The terms ``square'', ``cube'', ``hypercube'' come
from the intuitive representation 
of the corresponding functions 
by their tables of values.)
It is not difficult to see that every MDS code
 $M\subset \Sigma^{n+1}$ is the \emph{graph}
 $\{(f(\overline{x}),\overline{x}) \mid \overline{x}\in \Sigma^n\}$
 of some latin $n$-cube $f$ and vice versa, the graph of every latin $n$-cube is an MDS code of length $n$.
  We call a latin hypercube \emph{transitive} (\emph{topolinear})
if its graph is a transitive (topolinear) MDS code.
  Two latin hypercubes
are \emph{equivalent} (\emph{isotopic}) if their graphs are equivalent (isotopic, respectively).
  Clearly, two equivalent latin hypercubes
are or are not transitive (topolinear) simultaneously.
By the autotopism group $\mathrm{Ist}(f)$ of a latin hypercube we will mean the autotopism group of its graph.

It is easy to see that the operation $\star$ of every finite group $(\Sigma,\star)$ is a topolinear latin square.
If $f:\Sigma^2 \to\Sigma$ is a latin square and there exists an element $o \in \Sigma$
such that $f(x,o)=f(o,x)=x$ for every $x\in \Sigma$,
then the algebraic system $(\Sigma,f)$ is called a \emph{loop} and $o$ is called an \emph{identity element}.
Trivially, a loop has only one identity element;
if it is $0$, then the corresponding latin square is called \emph{reduced}.

Two loops $(\Sigma,f)$ and $(\Sigma,g)$ are called \emph{isotopic} to each other if
$f(x,y) = \varphi^{-1}g(\xi x, \psi y)$ for some isotopism $(\varphi,\xi, \psi)$;
if, additionally, $\varphi=\xi=\psi$, then the loops  $(\Sigma,f)$ and $(\Sigma,g)$ are called \emph{isomorphic}.
A loop $(\Sigma,f)$ is called a \emph{G-loop} if every loop isotopic to $(\Sigma,f)$ is isomorphic to $(\Sigma,f)$.

%=================================================================
\section{A connection between the isotopically transitive latin squares and the G-loops}\label{s:itG}
\begin{theorem}\label{th:IT=G}
A reduced latin square $f:\Sigma^2 \to\Sigma$ is isotopically transitive if and only if $(\Sigma,f)$ is a G-loop.
\end{theorem}
\begin{proof}
Let $(\Sigma,f)$ be a G-loop, where the identity element is $0$.
Consider arbitrary $a$, $b\in \Sigma$.
To find an autotopism that sends $(0,0,0)$ to $(f(a,b),a,b)$,
we will firstly choose an isotopism from $f$ to a reduced latin square $f''$
such that $(0,0,0)$ is mapped to $(f(a,b),a,b)$;
and then, we will apply an isomorphism from $f''$ to $f$.

Take arbitrary permutations $\xi$ and $\psi$ of $\Sigma$ that send
$0$ to $a$ and, respectively, to $b$;
and take an arbitrary permutation $\varphi$ that sends $f(a,b)$ to $0$.
To be definite, let $\xi$, $\psi$,  $\varphi$ be the transpositions that interchange $0$ and $a$,
$0$ and $b$, $f(a,b)$ and $0$, respectively.
Denote $f'(x,y)=\varphi f(\xi x, \psi y)$; 
define the permutations $\xi_0$ and $\psi_0$ as $\xi_0 x = f'( x,0)$ and $\psi_0 y = f'(0,y)$.
Then $f''(x,y) = f'(\xi_0^{-1}x, \psi_0^{-1}y)$ is a reduced latin square.
By the definition of a G-loop, we have $f(x,y) \equiv \tau^{-1}f''(\tau x, \tau y)$ for some $\tau$.
Since $\tau$ must send the identity element of $f$ to the identity element of $f''$, it fixes $0$.
Finally, $f(x,y)  \equiv  \tau^{-1}\varphi f(\xi\xi_0^{-1}\tau x, \psi\psi_0^{-1}\tau y)$, where
$\xi\xi_0^{-1}\tau 0= a$, $ \psi\psi_0^{-1}\tau 0 = b$. So, we have found an autotopism,
$(\varphi^{-1} \tau, \xi\xi_0^{-1}\tau,\psi\psi_0^{-1}\tau  )$, that sends $(0,0,0)$
to $(f(a,b),a,b)$; hence, $f$ is isotopically transitive.

Now assume that $f$ is an isotopically transitive reduced latin square.
Consider an arbitrary isotopism $\overline \tau$ such that $(\Sigma,\overline \tau f)$ is a loop.
Without loss of generality we may assume that its identity element is $0$
(otherwise we consider an isomorphic loop satisfying this condition).
Denote $(c=f(a,b),a,b)$ the preimage of $(0,0,0)$ under $\overline \tau$.
By the definition of the isotopical transitivity,
there is an autotopism $\overline\pi$ of $f$
that sends $(0,0,0)$ to $(c,a,b)$.
Then, $\overline\varphi=\overline\tau\overline\pi$ is an isotopism
of $f$ to $\overline \tau f$
that fixes $(0,0,0)$.
Since $0$ is an identity element of both $f$ and $\overline \tau f$,
such isotopism must be an isomorphism
(indeed, $y\equiv f(y,0)$ and $x\equiv \varphi_0^{-1}f(\varphi_1 x,0)$ imply $x\equiv \varphi_0^{-1} \varphi_1 x$;
i.e., $\varphi_1=\varphi_0$;
similarly, $\varphi_2=\varphi_0$).
Hence, $(\Sigma,f)$ is a G-loop.
\end{proof}

%Below, we will always assume that the identity element of a loop is $0$.
As shown in \cite{Wilson:74}, if $q$ is prime,
then every G-loop of prime order is a cyclic group;
a similar result for order $3p$, where $p>3$ is prime,
was established in \cite{Kunen:99}.
On the other hand, non-group G-loops are known to exist for all even orders larger than $5$
and all orders divisible by $p^2$ for some $p>2$ \cite{GR:82:Loops}.

%====================================================================================
\section{G-loops: examples}\label{s:G}
Let us consider some examples of groups and G-loops of order $2p$,
which will be used in the construction in the next section.
We set $\Sigma = \{0_0,1_0,\ldots,(p-1)_0,0_1,1_1,\ldots,(p-1)_1\}$.
Below, $+$ is the modulo $p$ addition, while $\oplus$ is the modulo $2$ addition.

\begin{enumerate}
\item The group $Z_p\times Z_2$ with the operation
$x_\zeta \bullet y_\xi=(x+y)_{(\zeta\oplus\xi)}$.

\item The \emph{dihedral} group $D_{2p}$ with the operation
$x_\zeta\circ y_\xi=((-1)^\xi x+y)_{(\zeta\oplus\xi)}$.

\item The loop $C_{2p}$ with the operation
$x_\zeta\ast y_\xi=((-1)^\xi x+y+\zeta\xi)_{(\zeta\oplus\xi)}$ \cite{Wilson:75}.
\end{enumerate}

\begin{figure}
$$
\begin{array}{c||c@{\,}c@{\,}c@{\,}c@{\,}c|c@{\,}c@{\,}c@{\,}c@{\,}c|}
\circ& 0_0& 1_0& 2_0& 3_0& 4_0& 0_1& 4_1& 3_1& 2_1& 1_1\\ \hline\hline
0_0& 0_0& 1_0& 2_0& 3_0& 4_0& 0_1& 1_1& 2_1& 3_1& 4_1\\
1_0& 1_0& 2_0& 3_0& 4_0& 0_0& 4_1& 0_1& 1_1& 2_1& 3_1\\
2_0& 2_0& 3_0& 4_0& 0_0& 1_0& 3_1& 4_1& 0_1& 1_1& 2_1\\
3_0& 3_0& 4_0& 0_0& 1_0& 2_0& 2_1& 3_1& 4_1& 0_1& 1_1\\
4_0& 4_0& 0_0& 1_0& 2_0& 3_0& 1_1& 2_1& 3_1& 4_1& 0_1\\ \hline
0_1& 0_1& 1_1& 2_1& 3_1& 4_1& 0_0& 1_0& 2_0& 3_0& 4_0\\
1_1& 1_1& 2_1& 3_1& 4_1& 0_1& 4_0& 0_0& 1_0& 2_0& 3_0\\
2_1& 2_1& 3_1& 4_1& 0_1& 1_1& 3_0& 4_0& 0_0& 1_0& 2_0\\
3_1& 3_1& 4_1& 0_1& 1_1& 2_1& 2_0& 3_0& 4_0& 0_0& 1_0\\
4_1& 4_1& 0_1& 1_1& 2_1& 3_1& 1_0& 2_0& 3_0& 4_0& 0_0\\ \hline
\end{array}\ \ \ \
\begin{array}{c||c@{\,}c@{\,}c@{\,}c@{\,}c|c@{\,}c@{\,}c@{\,}c@{\,}c|}
\ast& 0_0& 1_0& 2_0& 3_0& 4_0& 0_1& 4_1& 3_1& 2_1& 1_1\\ \hline\hline
0_0& 0_0& 1_0& 2_0& 3_0& 4_0& 0_1& 1_1& 2_1& 3_1& 4_1\\
1_0& 1_0& 2_0& 3_0& 4_0& 0_0& 4_1& 0_1& 1_1& 2_1& 3_1\\
2_0& 2_0& 3_0& 4_0& 0_0& 1_0& 3_1& 4_1& 0_1& 1_1& 2_1\\
3_0& 3_0& 4_0& 0_0& 1_0& 2_0& 2_1& 3_1& 4_1& 0_1& 1_1\\
4_0& 4_0& 0_0& 1_0& 2_0& 3_0& 1_1& 2_1& 3_1& 4_1& 0_1\\ \hline
0_1& 0_1& 1_1& 2_1& 3_1& 4_1& 1_0& 2_0& 3_0& 4_0& 0_0\\
1_1& 1_1& 2_1& 3_1& 4_1& 0_1& 0_0& 1_0& 2_0& 3_0& 4_0\\
2_1& 2_1& 3_1& 4_1& 0_1& 1_1& 4_0& 0_0& 1_0& 2_0& 3_0\\
3_1& 3_1& 4_1& 0_1& 1_1& 2_1& 3_0& 4_0& 0_0& 1_0& 2_0\\
4_1& 4_1& 0_1& 1_1& 2_1& 3_1& 2_0& 3_0& 4_0& 0_0& 1_0\\ \hline
\end{array}
$$
\caption{The value tables of the dihedral group $D_{10}=(\Sigma,\circ)$ and the G-loop $C_{10}=(\Sigma,\ast)$}
\end{figure}

\begin{proposition}\label{p:C2p} The loop $C_{2p}$, where $p$ is odd, is topolinear.
\end{proposition}
\begin{proof}
The transitivity is a direct consequence of Theorem~\ref{th:IT=G}
and the fact that $C_{2p}$ is a G-loop \cite{Wilson:75},
but we will give explicit formulas.
Consider four autotopisms of the MDS code
$M=\{(z_\psi, x_\zeta, y_\xi ) \ |\ z_\psi=x_\zeta\ast y_\xi\}$:

\begin{equation}\label{eq:a1-4}
\begin{array}{llll}
 \beta: & z_\eta\to (z-\frac12+\eta)_{\eta\oplus 1},& x_\zeta \to (-x)_\zeta, & y_\xi \to (y+\xi-\frac12)_{\xi\oplus 1}; \\
 \pi_a: & z_\eta\to (z+(-1)^\eta a)_\eta,  & x_\zeta \to (x+(-1)^\zeta a)_\zeta, & y_\xi \to y_\xi; \\
 \alpha: &z_\eta\to (-z+1-\eta)_{\eta}, & x_\zeta \to x_{\zeta\oplus 1}, & y_\xi \to
(-y)_{\xi\oplus 1}; \\
 \rho_b: & z_\eta\to (z+b)_\eta, & x_\zeta \to x_\zeta, & y_\xi \to (y+b)_\xi \\
\end{array}
\end{equation}
($\frac12$ is treated as $(p+1)/2$, modulo $p$). To prove that the transformations above are autotopisms,
it is sufficient to check the following:
if $(z_\eta,x_\zeta,y_\xi)$ satisfies $x_\zeta \ast y_\xi=z_\eta$ then
$\beta(z_\eta,x_\zeta,y_\xi)$, $\pi_a(z_\eta,x_\zeta,y_\xi)$, $\alpha(z_\eta,x_\zeta,y_\xi)$, $\rho_b(z_\eta,x_\zeta,y_\xi)$
also satisfy the similar equations.
For $\pi_a$ and $\rho_b$, the equations readily hold. For $\beta$, we have
\begin{eqnarray*}
 (-x)_\zeta \ast (y+\xi-{\textstyle\frac12})_{\xi\oplus 1}
&=& \left((-1)^{\xi\oplus 1}(-x)+y+\xi-{\textstyle\frac12} + \zeta(\xi\oplus 1)\right)_{\zeta\oplus\xi\oplus 1} \\
&=&
\left((-1)^\xi x +y +\zeta \xi -{\textstyle\frac12} +(\zeta\oplus \xi)\right)_{\zeta\oplus\xi\oplus 1} =
(z-{\textstyle\frac12}+\eta)_{\eta\oplus 1}
\end{eqnarray*}
(it is convenient to treat $\oplus$ using the identity $u\oplus v=u+v-2uv$). For $\alpha$, we have
\begin{eqnarray*}
 x_{\zeta\oplus 1} \ast (-y)_{\xi\oplus 1}
&=& \left((-1)^{\xi\oplus 1}x+(-y)+(\zeta\oplus 1)(\xi\oplus 1)\right)_{\zeta\oplus 1\oplus\xi\oplus 1} \\
&=&
\left(-(-1)^\xi x -y -\zeta \xi +1 -(\zeta\oplus \xi)\right)_{\zeta\oplus\xi} =
(-z+1-\eta)_{\eta}.
\end{eqnarray*}
It is easy to see that using the autotopisms from (\ref{eq:a1-4}) we can send
$(0_0,0_0,0_0)$ to any $(z'_{\eta'},x'_{\zeta'},y'_{\xi'})$
where $z'_{\eta'}=x'_{\zeta'}\ast y'_{\xi'}$:
\begin{itemize}
  \item if $\zeta'=1$, then we apply $\alpha$;
\item if $\zeta'\oplus\xi'=1$, then we apply $\beta$;
\item next, we apply $\pi_a$ to set the $x$-component (explicitly, $a=(-1)^{\zeta'}x'$);
 \item and finally, we apply $\rho_b$ to set $y$-component (explicitly, $b=y'-(\xi'-\frac12)(\zeta'\oplus \xi')$).
 \end{itemize}
So, $M$ is isotopically transitive.

For the topolinearity,
it remains to check that the order of the group
generated by the autotopisms $\beta$, $\pi_1$, $\alpha$, $\rho_1$
equals $|M|=(2p)^2$.
At first, we can see that the group generated by $\beta$ and $\pi_1$ is isomorphic to the Dihedral group $D_{2p}$
(see the action on the $x$-component).
Second, the group generated by $\alpha$ and $\rho_1$ is also isomorphic to the Dihedral group $D_{2p}$
(see the $y$-component). Next, each of $\beta$ and $\pi_1$ commutate with each of $\alpha$ and $\rho_1$.
We conclude that the total group is isomorphic to the direct product $D_{2p}\times D_{2p}$ of order $(2p)^2$.
\end{proof}

The group generated by the autotopisms $\alpha$, $\beta$, $\pi_1$, and $\rho_1$
(\ref{eq:a1-4}),
will be denoted by $IC_{2p}$.

\begin{corollary}\label{c:C2p}
There is a topolinear latin square that is not isotopic to a group operation.
\end{corollary}
\begin{proof}
 By Proposition~\ref{p:C2p}, the operation of the loop $C_{2p}$, $p$ odd,
 is an example of a topolinear latin square that is not a group operation.
 It remains to note that by Albert's theorem \cite[Theorem~2]{Albert:43},
 a loop that is not a group is not isotopic to a group.
\end{proof}

%====================================================================================
\section{Composition}\label{s:I}

In this section, 
we construct isotopically transitive latin hypercubes using composition 
of the loops $C_{2p}$ and $D_{2p}$, or $Z_p\times Z_2$ and $D_{2p}$. 
One of the two constructed classes provides an exponential (in $\sqrt{n}$) lower
bound on the number of nonequivalent topolinear latin $n$-cubes.

\subsection{Iterated groups}
If $(\Sigma,\star)$ is a group, then the system $(\Sigma,f)$ where
$f(x_1,\dots,x_n)=x_1\star\dots\star x_n$ is known as
an \emph{iterated group}.

\begin{proposition}\label{p:iter}
The operation of an iterated group is a topolinear latin hypercube.
%
% {\rm 2)} If a binary quasigroup $f$ is
% topolinear, %isotopically transitive,
% then an
% $n$-ary quasigroup obtained as iteration of $f$ is
% topolinear %isotopically transitive
% too.
\end{proposition}
\begin{proof}
Assume $(\Sigma,\star)$ is a group with the identity element $0$.
To each ${\bar y}=(y_0,y_1,\ldots,y_n)$ satisfying $y_0=y_1 \star \cdots \star y_n$,
we define the isotopism
$$\varphi_{\bar y}(x_0,x_1,\ldots,x_n) =
(
y_0 \star x_0,\
y_1 \star \cdots \star y_n \star x_1 \star y_n^{-1} \star \cdots \star y_2^{-1},\
 y_2 \star \cdots \star y_n \star x_2 \star y_n^{-1} \star \cdots \star y_3^{-1},\
 \ldots,\
   y_n \star x_n).$$
It is not difficult to check that 1) $\varphi_{\bar y}$ is an autotopism of the iterated group;
2) $\varphi_{\bar y}$ sends $\overline 0$ to $\bar y$;
3) $\varphi_{\bar y}\varphi_{\bar z} = \varphi_{\varphi_{\bar y}(\bar z)}$.
\end{proof}

\begin{proposition}\label{p:id-iterated}
Let $h$ be the operation of an $n$-ary iterated group. Then
  for every  $\overline{b}\in \Sigma^n$ satisfying $h(\overline{b})=0$,
there exists an autotopism $\overline{\theta}=(\mathrm{Id},\theta_1,\ldots,\theta_n)$ of $h$ such that
$\overline{\theta}(0,\overline{b})=\overline 0$.
\end{proposition}
\begin{proof}
Take
$\overline{\theta} = \varphi_{(0,\bar b)}^{-1}$, where $\varphi_{(0,\bar b)}$ is defined in the proof of Proposition~\ref{p:iter}.
\end{proof}

 Consider a group $(\Sigma,\star)$.
 Define $M(\star)=\{(x_0,x_1,x_2) : x_0=x_1\star x_2\}$
and $A^i(\star)=\{\tau_i \ |\ (\tau_0,\tau_1,\tau_2)\in {\rm
Ist}(M(\star))\mbox{ for some }\tau_{i+1\bmod 3}, \tau_{i+2\bmod 3}\}$, $i=0,1,2$.

\begin{proposition}\label{prodop11}
 For a group $(\Sigma,\star)$, the set $A^i(\star)$ does not depend on $i$ and forms a subgroup of the group of permutations of $\Sigma$.
\end{proposition}
\begin{proof}
 If $(\tau_0,\tau_1,\tau_2)$, $(\pi_0,\pi_1,\pi_2)\in {\rm
Ist}(M(\star))$, then $\tau_0\pi_0\in A^0(\star)$. Hence,
$A^0(\star)$ is a subgroup of the group of permutations of $\Sigma$.

Assume that $(\tau_0,\tau_1,\tau_2)\in \mathrm{Ist}(\star)$.
Let us show that $(\tau_1,\tau_0,\iota \tau_2 \iota)\in \mathrm{Ist}(\star)$,
where $\iota$ is the permutation that interchange the elements with their inverses.
Consider the following sequence of equivalent equations.
We start with
$$\tau_1 z =\tau_0  x \star \iota \tau_2 \iota y.$$
Multiplying the both parts by $(\iota \tau_2 \iota y)^{-1}=\tau_2 \iota y$ on the right, we have
$$ \tau_1 z \star \tau_2 \iota y = \tau_0  x. $$
Since $(\tau_0,\tau_1,\tau_2)$ is an autotopism of $\star$, the last is equivalent to
$$ z \star  \iota y =  x. $$
 Then, multiplying by $y=(\iota y)^{-1}$, we get
 $$ z  =  x\star  y. $$
 We see that $(\tau_1, \tau_0, \iota \tau_2 \iota)$ also belongs to $\mathrm{Ist}(\star)$,
 which means that $A^0(\star)=A^1(\star)$. Similarly, $A^0(\star)=A^2(\star)$.
\end{proof}

\begin{proposition} \label{prodop12}
Let $(\Sigma,h)$ be the iterated group of $(\Sigma,\star)$.
Let $\sigma \in A^0(\star)$.
Then there is an autotopism
$(\sigma,\overline{\tau})$ of $h$.
\end{proposition}

 \begin{proof} We proceed by induction on the number $n$ of arguments of $h=h_n$.
 For $n=2$, the statement is straightforward from the definition of $A^0(\star)$.
If $n>2$, then
\begin{equation}\label{eq:hn}
 \sigma h_n(x_1,\dots,x_n)\equiv \sigma( h_{n-1}(x_1,\dots,x_{n-1})\star x_n)
\equiv \pi h_{n-1}(x_1,\dots,x_{n-1})\star \tau_n x_n
\end{equation}
for some permutations $\pi$ and $\tau_n$ (whose existence follows from the definition of $A^0(\star)$).
By Proposition~\ref{prodop11}, we have $\pi\in A^0(\star)$.
Then, by the induction hypothesis,
\begin{equation}\label{eq:hn-1}
\pi h_{n-1}(x_1,\dots,x_{n-1})\equiv h_{n-1}(\tau_1 x_1,\dots,\tau_{n-1}x_{n-1})
\end{equation}
for some $\tau_1$, \ldots, $\tau_{n-1}$. Substituting (\ref{eq:hn-1}) into (\ref{eq:hn}),
we find the required autotopism.
\end{proof}

\subsection{Composition of type $f(h,\ldots,h)$}
The following lemma generalizes \cite[Proposition~8]{Pot:trans}.

\begin{lemma}\label{l:gen}
Let $f$ be a latin $n$-cube, and let $G_f\le \mathrm{Ist}(f)$ be a autotopism group of $f$ that acts transitively on its graph.
%%%%(so, $f$ is isotopically transitive).
Let for every $i$ from $1$ to $n$,  $h_i$ be the $m_i$-ary operation of an iterated group $(\Sigma,\star)$.
Let %%%%Assume that 
for every $(\sigma_0,\sigma_1,\ldots,\sigma_n)$ from $G_f$,
all $\sigma_i$, $i=1,\ldots,n$, lie in  $A^0(\star)$.
Then the $m$-ary quasigroup $f(h_1(z_1),\dots,  h_n(z_n))$, where $m
= m_1 + \dots +m_n$, is isotopically transitive.
\end{lemma}
\begin{proof}
Consider an arbitrary tuple
$b_0,\overline{b}_1,\dots,\overline{b}_n$ satisfying
$f(h_1(\overline{b}_1),\dots,h_n(\overline{b}_n))=b_0$. There exists
an isotopism $\sigma\in G_f$ such that $\sigma_0b_0=0$ and $\sigma_i
h_i(\overline{b}_i)=0$ for all $i\in\{1,\dots,n\}$. By
Proposition~\ref{prodop12} there exist $\overline{\tau}_1,
\dots,\overline{\tau}_n$ such that
$$\sigma_0f(h_1(\overline{\tau}_1\overline{z_1}),
\dots,h_n(\overline{\tau}_n\overline{z_n}))\equiv
\sigma_0f(\sigma_1h_1(\overline{z_1}),\dots,\sigma_nh_n(\overline{z_n}))\equiv
f(h_1(\overline{z_1}),\dots,h_n(\overline{z_n})).$$ 
So, we have
an isotopism that sends 
$(b_0,\overline{b}_1,\dots,\overline{b}_n)$ 
to
$(0,\overline{\tau}_1\overline{b}_1,\dots,\overline{\tau}_n\overline{b}_n)$.

By Proposition~\ref{p:id-iterated}, for every $i\in\{1,\dots,n\}$ there
exists an isotopism $\overline{\theta_i}$ such that
$\overline{\theta_i}\overline{\tau}_i\overline{b}_i=\overline{0}$
and $h_i(\overline{\theta_i}\overline{z_i})\equiv
h_i(\overline{z_i})$. So, we have got an isotopism that sends
$(0,\overline{\tau}_1\overline{b}_1,\dots,\overline{\tau}_n\overline{b}_n)$
 to
$\overline{0}$.
\end{proof}

The next proposition shows that the roles of $f$ and $h_i$s 
in Lemma~\ref{l:gen} can be played by the loop $C_{2p}$ and iterations of the dihedral group, respectively.
\begin{proposition}\label{p:IC2p-in-A0(D2p)}
 For every $(\sigma_0,\sigma_1,\sigma_2)\in IC_{2p}$, the permutations
 $\sigma_0$, $\sigma_1$, and $\sigma_2$ belong to $A^0(D_{2p})$,
 where $D_{2p}=(\Sigma,\circ)$ is the dihedral group,
 $u_\varepsilon\circ v_\kappa=((-1)^\kappa u+v)_{\varepsilon\oplus\kappa}$.
\end{proposition}

\begin{proof}
We have to prove that for all $(\sigma_0,\sigma_1,\sigma_2)\in IC_{2p}$ and for
$\sigma=\sigma_0$, $\sigma=\sigma_1$, or $\sigma=\sigma_2$, there are
$\tau_1$ and $\tau_2$ such that $\sigma(u_\varepsilon\circ v_\kappa)=\tau_1 u_\varepsilon\circ\tau_2 v_\kappa$.
It is sufficient to prove this for the generators
(\ref{eq:a1-4}) of the group $IC_{2p}$. In the table, we suggest $\tau_1$ and $\tau_2$
for each possible $\sigma$ that comes from~(\ref{eq:a1-4}).
$$
\begin{array}{|c|c|c|c|} \hline
\sigma w_\omega= \hfill& \tau_1u_\varepsilon=\hfill & \tau_2v_\kappa= \hfill & \sigma(u_\varepsilon \circ v_\kappa)=\tau_1 u_\varepsilon \circ\tau_2 v_\kappa \\ \hline\hline
w_{\omega\oplus 1} & u_{\varepsilon\oplus 1} & v_\kappa &
                 \begin{array}{@{}r@{}}
                  \left((-1)^\kappa u+v\right)_{\varepsilon\oplus\kappa\oplus1}\\=\left((-1)^\kappa u+v\right)_{\varepsilon\oplus1\oplus\kappa}
                 \end{array}\\ \hline
(-w)_{\omega\phantom{}} & (-u)_{\varepsilon\phantom{}} & (-v)_\kappa &
                   \begin{array}{@{}r@{}}\left(-((-1)^\kappa u+v)\right)_{\varepsilon\oplus\kappa}\\=\left((-1)^\kappa (-u)-v\right)_{\varepsilon\oplus\kappa}
                   \end{array}\\ \hline
(-w)_{\omega\oplus 1} & (-u)_{\varepsilon\oplus 1} & (-v)_\kappa &
            \begin{array}{@{}r@{}}\left(-((-1)^\kappa u+v)\right)_{\varepsilon\oplus\kappa\oplus 1}\\=\left((-1)^\kappa (-u)-v\right)_{\varepsilon\oplus 1\oplus\kappa}
            \end{array}
\\ \hline
 (w+(-1)^\omega a)_{\omega\phantom{}} &  (u+(-1)^\varepsilon a)_{\varepsilon\phantom{}} & v_\kappa &
            \begin{array}{@{}r@{}}\left((-1)^\kappa u+v+(-1)^{\varepsilon\oplus\kappa}a\right)_{\varepsilon\oplus\kappa}\\=\left((-1)^\kappa (u+(-1)^\varepsilon a)+v\right)_{\varepsilon\oplus\kappa}
            \end{array}
\\ \hline
 (w+b)_{\omega\phantom{}} & u_{\varepsilon\phantom{}} &(v+b)_\kappa &
            \begin{array}{@{}r@{}}\left((-1)^\kappa u+v+b\right)_{\varepsilon\oplus\kappa}\\=\left((-1)^\kappa u+v+b\right)_{\varepsilon\oplus\kappa}
            \end{array}
\\ \hline
 (-w+1-\omega)_{\omega\phantom{}}  & (-u-\varepsilon)_{\varepsilon\phantom{}} & (-v-\kappa+1)_\kappa &
       \begin{array}{@{}r@{}}
        \left(-(-1)^\kappa u-v+1-\varepsilon\oplus\kappa\right)_{\varepsilon\oplus\kappa}\\=\left((-1)^\kappa (-u-\varepsilon)-v-\kappa+1\right)_{\varepsilon\oplus\kappa}
       \end{array}
\\ \hline
 (w-\frac12+\omega)_{\omega\oplus 1}   &  (u+\varepsilon)_{\varepsilon\oplus1} & (v+\kappa-\frac12)_\kappa &
 \begin{array}{@{}r@{}}
      \left((-1)^\kappa u+v-\frac12+\varepsilon\oplus\kappa\right)_{\varepsilon\oplus\kappa\oplus1} \\
       \left((-1)^\kappa (u+\varepsilon)+v+\kappa-\frac12\right)_{\varepsilon\oplus1\oplus\kappa}
 \end{array}
 \\ \hline
\end{array}
$$
In each case, the identity $\sigma(u_\varepsilon\circ v_\kappa)=\tau_1 u_\varepsilon\circ\tau_2 v_\kappa$ is straightforward to check, see the right column
(in the last two cases, the identity $\zeta\oplus\xi=(-1)^\xi\zeta+\xi$ is useful).
\end{proof}

\subsection{Explicit constructions: $f$ is $C_{2p}$ or $Z_p\times Z_2$, $h$ is iterated $D_{2p}$}
Now, we are ready to construct two series of isotopically transitive latin hypercubes,
which are nonequivalent to an iterated group.
\begin{theorem}\label{th:iter1}
Let $\ast$ be  the operation of the loop $C_{2p}$, defined in Section~\ref{s:G}.
Let $h_i$, $i=1,2$,  be the $m_i$-ary operation of the iterated dihedral group $D_{2p}$.
Then the latin hypercube $g$, $g=h_1(\overline{z_1})\ast h_2(\overline{z_2})$, is
isotopically transitive.
\end{theorem}
\begin{proof}
The hypothesis of Lemma~\ref{l:gen} is
satisfied by Propositions~\ref{p:C2p}
and~\ref{p:IC2p-in-A0(D2p)} with $G_f=IC_{2p}$.
\end{proof}

In the next theorem, 
we also use composition 
to construct isotopically transitive latin hypercubes.
Moreover, %%%%we can say that 
the resulting objects 
are %%%%will be 
topolinear. 
The following statement is convenient
in establishing the topolinearity.

\begin{proposition}\label{p:topolin-from-0fixator}
Let $M\subseteq \Sigma^n$ be 
an isotopically transitive MDS code.
Let $G<\mathrm{Ist}(M)$ be an autotopism group 
that acts transitively on $M$,
and let for some $i$ from $0$ to $n-1$
every $\overline{\sigma}=(\sigma_0,\ldots,\sigma_{n-1})$ from $G$
such that $\overline{\sigma}(\overline{0})=\overline{0}$
satisfy $\sigma_i=\mathrm{id}$.
Then $M$ is topolinear with regular group $G$.
\end{proposition}

\begin{proof}
Assume that $\overline\sigma(\overline{0})=\overline{0}$;
by the hypothesis, $\sigma_i=\mathrm{id}$.
Assume without loss of generality that $i=0$.
Let us show that $\sigma_1=\mathrm{id}$.
For every $b\in \Sigma$, there exists
$a\in \Sigma$ such that $\overline b=(a,b,0\dots,0)\in M$.
Then,
$\sigma(\overline b)=(a,\sigma_1(b),0\dots,0)\in M$.
By the definition of an MDS code, we have $\sigma_1(b)=b$.
Similarly, we have $\sigma_j=\mathrm{id}$ for all $j$ from $1$ to $n-1$.
Hence, the only isotopism that fixes $\overline{0}$ is the identity isotopism.
It follows that the subgroup $G$ is regular.
\end{proof}

\begin{theorem}\label{th:iter2}
Let $f$ be the $n$-ary operation of the iterated group  $Z_p\times Z_2=(\Sigma,\bullet)$.
Let $h_i$, $i\in \{1,\dots,n\}$, be the $m_i$-ary operation of the iterated dihedral group $D_{2p}$.
Then the latin hypercube $g$, $g=f(h_1(\overline{z_1}),\ldots, h_n(\overline{z_n}))$, is topolinear.
%
% {\color{red} [??? ³-T³³³³³- ³¿³¿ ³-³-³- ³³³  ³³³-³+?] $M=\{(x,\overline{z_1},\dots,\overline{z_n})\ |\
% x=f(h_1(\overline{z_1}),\dots,h_n(\overline{z_n}))\}$ is topolinear.}
\end{theorem}
\begin{proof}
By Proposition~\ref{p:iter}, $h_1$, \ldots, $h_n$, and $f$ are topolinear latin hypercubes.
Then,
the hypothesis of  Lemma~\ref{l:gen}
is satisfied with $G_f$ being the translation group
$\{ \bar\sigma_{(b_0,b_1,...,b_n)} \mid b_0 = b_1 \bullet \cdots \bullet b_n,\  \bar\sigma_{(b_0,b_1,...,b_n)} (v_0,v_1,\ldots,v_n) =
(b_0\bullet v_0,b_1 \bullet v_1,\ldots,b_n \bullet v_n) \}$
(indeed, any permutation $\sigma$ of form $\sigma(v_i)=b_i \bullet v_i$ is a combination
of the permutations considered in the proof of Proposition~\ref{p:IC2p-in-A0(D2p)}, see the first and the fifth rows of the table).
Hence, $g$ is isotopically transitive.

Consider the autotopism group $G$ obtained accordingly to Lemma~\ref{l:gen} and acting transitively on the graph $M$ of $g$.
For every $(\sigma_0,\sigma_1,\ldots)\in G$, we have $\sigma_0(v_0)= b_0\bullet  v_0$.
So, the conjecture of Propositions~\ref{p:topolin-from-0fixator} is satisfied for the first coordinate of $M$
(which corresponds to the value of $g$).
Hence, $g$ is topolinear.
\end{proof}

\begin{corollary}\label{cor:No} For every integer $p\geq 2$, the number of mutually nonequivalent topolinear
latin $N$-cubes of order $2p$ grows at least as
$\frac{1}{4N\sqrt{3}}e^{\pi\sqrt{2N/3}}(1+o(1))$.
\end{corollary}
\begin{proof}
As follows from the theorem on canonical decomposition of an $n$-ary
quasigroup \cite{Cher}, two MDS codes $M$, $M'$ obtained from
different sets $\{m_1,\ldots,m_n\}$, $\{m'_1,\ldots,m'_{n'}\}$ as in
Theorem~\ref{th:iter2} are nonequivalent. So, (permutably)
nonequivalent partitions $N=m_1+\ldots+m_n$ lead to nonequivalent
codes. The number of nonequivalent partitions is known to be
$\frac{1}{4N\sqrt{3}}e^{\pi\sqrt{2N/3}}(1+o(1))$
\cite{Andrews:partitions}.
\end{proof}

%==================================================================================
\section{Topolinear latin hypercubes from quadratic functions}\label{s:Q}
In this section, we consider the construction of 
topolinear latin hypercubes based on quadratic functions.
The lower bound (Corollary~\ref{cor:No2}) 
on the number of topolinear latin hypercubes
based on this construction
is close to the known upper bound \cite[Theorem~2]{KroPot:2012:propel}.

Let $q=p^k$, where $p$ is prime.
We will assume that $\Sigma_q$ is equipped with the structure of the field $GF(p^k)$ and
that $\Sigma_{q^2}=\Sigma_q\times \Sigma_q$ consists of the pairs $[a,b]$ of elements of $\Sigma_q$.

\begin{theorem}\label{th:quadratic-topolinear}
Assume that the latin $n$-cube $f$ over
$\Sigma_{q^2}^n$
is defined by
$$
f\left([x_1,y_1],\dots,[x_n,y_n]^{\vphantom{1}}\right) = \Bigl[-\sum_{i=1}^nx_i,\
-\sum_{i=1}^ny_i-r(x_1,\ldots,x_n)\Bigr],
$$
where
\begin{equation}\label{eq:r}
  r(x_1,\ldots,x_n)=\sum_{i,j=1}^n \alpha_{ij}x_ix_j %%%bbb +\sum_{i=1}^n\beta_i(x_i)
\end{equation}
with
%%%bbb $\beta_i:\Sigma_q\rightarrow \Sigma_q$ being arbitrary functions and
$\alpha_{ij}$ being some constants, $i,j\in\{1,\ldots,n\}$.
Then $f$ is a topolinear latin hypercube.
\end{theorem}
\begin{proof}
We will firstly prove that the graph $M$ of $f$ is an MDS code.
It is sufficient to show that arbitrarily fixing
$n-1$ coordinates uniquely defines the remaining
coordinate $[x_i,y_i]$ of a vertex from $M$.
Indeed, 
the formula for the first component of $f$ defines the remaining $x_i$;
the formula for the second component defines $y_i$.

%%%bbb Without loss of generality we assume that $\overline{0}\in M$ and $\beta_i(0)=0$ for every $i\in \{1,\dots,n\}$.
Now establish the isotopical transitivity of $M$.
Let
$([a_0,b_0],\dots,[a_n,b_n])\in M$.
Then the isotopism
\begin{eqnarray*}
\left([x_0,y_0],\ldots,[x_n,y_n]^{\vphantom{1}}\right) &\to&
\left([\sigma_0 (x_0),\tau_0 (y_0)],\ldots,[\sigma_n (x_n),\tau_n (y_n)]^{\vphantom{1}}\right):
\\
\sigma_i(x_i)&=&x_i-a_i,
\\
\tau_i(y_i)&=&y_i+x_i\sum_{j=1}^n \alpha_{ij}a_j+ x_i\sum_{j=1}^n
\alpha_{ji}a_j %%%bbb -\beta_i(x_i-a_i)+\beta_i(x_i)
-\sum_{j=1}^n \alpha_{ij}a_ja_i
\end{eqnarray*}
(we define $\alpha_{0j}=0$)
belongs to the group
$\mathrm{Ist}(M)$
and sends
$([a_0,b_0],\dots,[a_n,b_n])\in M$
to
$([0,\tau_0b_0],\dots,[0,\tau_nb_n])\in M$.
The isotopism
$\sigma'_i(x_i)=x_i$, $\tau'_i(y_i)=y_i-c_i$
is contained in
$\mathrm{Ist}(M)$
and sends
$([0,c_1],\dots,[0,c_n])\in M$
to
$([0,0],\dots,[0,0])$.
By Propositions~\ref{p:topolin-from-0fixator} and~\ref{p:id-iterated},
 $M$ is topolinear.
\end{proof}

\begin{corollary}\label{cor:No2}
Let $q=p^k$,
where $p$ is prime, and let $s\geq 1$.
Then there are at least $q^{\frac{n^2}2(1+o(1))}$ nonequivalent
topolinear latin $n$-cubes of order $q^2 s$.
\end{corollary}
\begin{proof}
At first, consider the case $s=1$.
Choosing different coefficients $\alpha_{i,j}$, $1\le i\le j\le n$,
we obtain $q^{\frac{n^2}2(1+o(1))}$ different functions $r$, see (\ref{eq:r}),
and, by Theorem~\ref{th:quadratic-topolinear},
the same number of different topolinear latin $n$-cubes. Dividing this number by the number
$(q^2!)^{n+1} (n+1)! = q^{O(n\ln n)}$ of different isometries of $\Sigma_{q^2}^{n+1}$,
we get the lower bound $q^{\frac{n^2}2(1+o(1))}$
on the number of nonequivalent topolinear latin $n$-cubes.

As follows from Proposition~\ref{p:cartesian-product},
from each topolinear MDS code $M$ in $\Sigma_{q^2}^{n}$,
we can construct a topolinear MDS code
$M\times S$ in $\Sigma_{q^2 s}^{n}$, where $S$ is the graph of an iterated group of order $s$.
This argument expands the result to arbitrary $s$.
\end{proof}

In \cite{KroPot:2012:propel}, the upper bounds $\exp(O(n^2 \ln^2 n))$ and  $\exp(O(n^2 \ln n))$
on the number of transitive and, respectively, propelinear codes were derived.
As we see, by the order of the logarithm, 
the lower bound from Corollary~\ref{cor:No2} 
is rather close to the known upper bounds.

%----------------------------------------------------------------------------------------
\section{A characterization of the set of isotopically transitive latin hypercubes of order $4$}\label{s:4}
Now consider the case $q=2$. 
As we will see in this section, in this case, 
the construction in Section~\ref{s:Q} exhausts the class of the isotopically transitive latin hypercubes.
\subsection{Semilinear MDS codes}
  A latin $n$-cube over
  $\Sigma_{2^2}$ (as well as the corresponding MDS code in $\Sigma_{2^2}^{n+1}$) is called \emph{semilinear}
  if it is isotopic to a latin $n$-cube $f$
defined by
\begin{equation}\label{eq:1}
f\left([x_1,y_1],\dots,[x_n,y_n]^{\vphantom{1}}\right) = \Bigl[\sum_{i=1}^nx_i,\
\sum_{i=1}^ny_i+r(x_1,\ldots,x_n)\Bigr]
\end{equation}
 with an arbitrary Boolean function $r$ (here and in what follows, $+$ means the addition modulo~$2$).
A semilinear latin $n$-cube (and its graph) is called \emph{linear} 
if it is isotopic to the latin $n$-cube (\ref{eq:1}) with
$r(\ldots)\equiv 0$. 
It is easy to see that every subcode 
of a semilinear (linear) MDS code 
is semilinear (linear, respectively).

\begin{proposition}%
[{\cite[Theorem 8.8]{Zaslavsky:2012}}]%
\label{p:linear-from-3subcodes}
If all $4$-subcodes of an MDS code 
$M\subset \Sigma_{2^2}^{n+1}$ 
($n\geq 3$) are linear,
then $M$ is linear itself.
\end{proposition}
% \begin{proof}
% The ``only if'' part is trivial.
% For $n=4$, the ``if'' claim can be checked directly, utilizing the characterization of all $5$ nonequivalent codes \cite{ZZ:2002},
% which can be found below. Then, it is sufficient to convince that any MDS code whose $4$-subcodes are all linear is linear itself.
% This was proved in \cite[Theorem 8.8]{Zaslavsky:2012}.
% \end{proof}

After one definition and auxiliary Propositions~\ref{p:two-double-codes}--\ref{p:non-transitive-length4-MDS}, 
we will prove a similar statement 
for the semilinear MDS codes (Lemma~\ref{l:from-n-1-to-n}),
which plays a crucial role in the proof of the main result of this section (Theorem~\ref{th:order4}).
% The similar statement for the semilinear MDS codes does not hold.
% However, as we well see below, the weakened variant of the statement,
% where the $4$-subcodes are considered instead of the $3$-subcodes,
% will be valid for the semilinear MDS codes.

\subsection{Linearizations}
A set $D\subset \Sigma_{2^2}^{n+1}$
is called a \emph{linearization} with components
$D_0$, \ldots, $D_n$
if its characteristic function $\chi_D$ 
can be represented as
\begin{equation}\label{eq:11}
 \chi_D(z_0,\ldots,z_n)=\chi_{D_0}(z_0)+\dots+\chi_{D_n}(z_n) + \gamma
\end{equation}
  where $D_i$ ($i\in \{0,\ldots,n\}$) 
  are $2$-subsets of $\{[0,1],[1,0],[1,1]\}$
  and $\gamma\in \{0,1\}$ is a constant. 
  For each linearization, 
  the representation (\ref{eq:11}) is unique.

\begin{proposition}\label{p:two-double-codes}
Assume that an MDS code $M\subset \Sigma_{2^2}^{n+1}$ 
is a subset of two different linearizations $D$ and $E$,
with components
$D_0$, \ldots, $D_n$ and $E_0$, \ldots, $E_n$,
respectively.
Then $D_i\ne E_i$ for every $i$ from $0$ to  $n$.
\end{proposition}
\begin{proof}
Since $D$ and $E$ are different, we have $|D_j\cap E_j|=1$ for some $j$.
Seeking a contradiction, assume $D_i = E_i$ for some $i$. 
Without loss of generality, $i=0$ and $j=1$. 
The two linearizations 
\begin{eqnarray*}
  D^{0,1} &=& \{ (z_0,z_1) \mid (z_0,z_1,[0,0],\ldots,[0,0]) \in D \}, \\
 E^{0,1} &=& \{ (z_0,z_1) \mid (z_0,z_1,[0,0],\ldots,[0,0]) \in E \}
\end{eqnarray*}
include the same $2$-subcode of $M$. 
But it is straightforward that $D^{0,1} \cap E^{0,1} $ has no MDS-code subsets:
$$
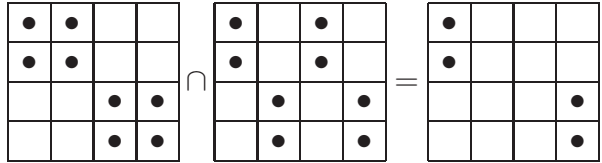
\begin{array}{|c|c|c|c|}
\hline
 \bullet & \bullet &         &         \\ \hline
 \bullet & \bullet &         &         \\ \hline
         &         & \bullet & \bullet \\ \hline
         &         & \bullet & \bullet \\ \hline
\end{array}
\cap
\begin{array}{|c|c|c|c|}
\hline
 \bullet &         & \bullet &         \\ \hline
 \bullet &         & \bullet &         \\ \hline
         & \bullet &         & \bullet \\ \hline
         & \bullet &         & \bullet \\ \hline
\end{array}
=
\begin{array}{|c|c|c|c|}
\hline
 \bullet & \phantom{\bullet} & \phantom{\bullet} &         \\ \hline
 \bullet &         &         &         \\ \hline
         &         &         & \bullet \\ \hline
         &         &         & \bullet \\ \hline
\end{array}
$$
We have found a contradiction. Hence, $D_i\ne E_i$ for all $i$ from $0$ to  $n$.
\end{proof}

\begin{proposition}\label{p:(semi)linear-from-2MDS}
{\rm 1) \cite[Proposition~2.6]{KroPot:4}}
An MDS code $M\subset \Sigma_{2^2}^{n+1}$
is semilinear if and only if
there exists a linearization
$D\subset \Sigma_{2^2}^{n+1}$ such that
$M\subset D$.

{\rm 2) \cite[Assertion~15(d)]{Pot:2012:partial}}
A semilinear MDS code
$M\subset \Sigma_{2^2}^{n+1}$
is not linear
if and only if
such a linearization $D$ is unique.
\end{proposition}
\begin{proof}
  1) ``Only if''. If $f$ is defined by (\ref{eq:1}), 
  then its graph $M$ is a subset of the linearization 
  \begin{equation}\label{eq:Dx}
   D=\{([x_0,y_0],\ldots,[x_n,y_n])\mid x_0+\dots +x_n = 0\}. 
  \end{equation}
  Every MDS code isotopic to $M$ is a subset of a linearization
  isotopic to $D$.
  
  ``If''. Assume without loss of generality that $M\subset D$, where $D$ is from (\ref{eq:Dx}).
  For every $(x_0,\ldots,x_n)$ satisfying $x_0+\dots +x_n = 0$, the set 
  $\{(y_0,\ldots,y_n) \mid ([x_0,y_0],\ldots,[x_n,y_n])\in M\}$ is an MDS code in $\Sigma_2^{n+1}$.
  It is one of $\{(y_0,\ldots,y_n) \mid y_0+\dots+y_n=0\}$, 
  $\{(y_0,\ldots,y_n) \mid y_0+\dots+y_n=1\}$. In the former case we define $r(x_1,\ldots,x_n)=0$; in the latter, $r(x_1,\ldots,x_n)=1$.
  Then, $M$ is the graph of (\ref{eq:1}); consequently, it is semilinear.
  
2) ``Only if''. Assume that an MDS code $M$ 
is a subset of two different linearizations $D$ and $E$, 
with components $D_0$, \ldots, $D_n$ 
and $E_0$, \ldots, $E_n$, 
respectively. 
By Proposition~\ref{p:two-double-codes}, 
$|D_i\cap E_i|=1$ for each $i$. 
Considering an appropriate isotopism, 
we can assume without loss of generality that 
$D_i=\{[1,0],[1,1]\}$ and $E_i=\{[0,1],[1,1]\}$.
Also, we can assume that the constant $\gamma$
in (\ref{eq:11}) is equal to $1$ both for $D$ and $E$.
Then, $D$ satisfies (\ref{eq:Dx})
and $E$ satisfies
  \begin{equation}\label{eq:Ex}
   E=\{([x_0,y_0],\ldots,[x_n,y_n])\mid y_0+\dots +y_n = 0\}. 
  \end{equation}
Now, it is straightforward that 
$D \cap E$ is the graph of (\ref{eq:1}) 
with $r(x_1,\ldots,x_n)\equiv 0$.
Hence $M=D \cap E$ is linear by the definition.

``If''. If $M$ is the graph of the
latin $n$-cube (\ref{eq:1}) with
$r(\ldots)\equiv 0$, 
then it is a subsets of two linearizations 
$D$ (\ref{eq:Dx}) and $E$ (\ref{eq:Ex}).
Every MDS code isotopic to $M$ is a subset of two linearizations
isotopic to $D$ and $E$.
\end{proof}

\subsection{MDS codes of length $4$}

 The
paper \cite{KroPot:4} contains a characterization of the MDS codes
in $\Sigma_4^n$. In particular, the following is true \cite{ZZ:2002}.

\begin{proposition}\label{p:len4}
There are $5$ equivalence classes
of MDS codes of length $4$.
Four of them are semilinear,
with representatives corresponding to the Boolean functions
\begin{eqnarray}
r_1(x_1,x_2,x_3)&\equiv& 0, \nonumber \\
r_2(x_1,x_2,x_3)&\equiv&x_1x_3 + x_2x_3, \nonumber \\
r_3(x_1,x_2,x_3)&\equiv&x_1x_2+ x_1x_3 + x_2x_3,  \nonumber \\
r_4(x_1,x_2,x_3)&\equiv&(x_1+1)x_2x_3. \label{eq:r4}
\end{eqnarray}
Any non-semilinear MDS code in $\Sigma_{2^2}^4$ is equivalent to
\begin{equation}\label{eq:H}
H=\{(z_0,z_1,z_2,z_3) \ |\ z_0 =z_1 \star (z_2 \diamond z_3)\},
\end{equation}
where $\star$ and $\diamond$ are group operations isotopic to $Z_4$,
with the same identity element but different elements of order $2$.
\end{proposition}

The following two propositions can be checked directly.

\begin{proposition}\label{p:3subcodes-of-nonlinearMDS}
% {\rm 1)} If an MDS code $M\subset \Sigma_{2^2}^4$ is nonlinear then some two (or
% four) parallel (obtained by fixing the same
% argument by different values) $3$-subcodes of $M$ are nonlinear.
% {\rm 2)}
If $M\subset \Sigma_{2^2}^4$ is one of the three semilinear MDS codes corresponding to the functions $r_2$, $r_3$, $r_4$ of (\ref{eq:r4}).
Then the $3$-subcodes
 \begin{eqnarray*}
 T^a&=&\{(z_0,z_2,z_3) \mid (z_0,a,z_2,z_3)\in M\}, \\
 T^b&=&\{(z_0,z_2,z_3) \mid (z_0,b,z_2,z_3)\in M\}, \\
 T^c&=&\{(z_0,z_1,z_3) \mid (z_0,z_1,c,z_3)\in M\},
 \end{eqnarray*}
where $a=[0,0]$, $b=[0,1]$, $c=[1,0]$ are nonlinear; moreover, $T^a \cup T^b$ is a linearization.
%
% exist two nonlinear parallel $3$-subcodes of $M$
% whose union is a linearization.
\end{proposition}

\begin{proposition}\label{p:non-transitive-length4-MDS}
The non-semilinear MDS code $H$, see (\ref{eq:H}), and the semilinear MDS code $S_4$
with the function $r_4$, see (\ref{eq:r4}), are not isotopically transitive.
\end{proposition}

\subsection{The crucial lemma}
\begin{lemma}\label{l:from-n-1-to-n}
If every $n$-subcode of an MDS code
$M\subset \Sigma_{2^2}^{n+1}$, $n\geq 4$, is semilinear,
then $M$ is semilinear too.
\end{lemma}
\begin{proof}
 If $M$ does not have nonlinear $4$-subcodes, then  $M$ is linear by
Proposition~\ref{p:linear-from-3subcodes}. In the other case, by
Proposition~\ref{p:3subcodes-of-nonlinearMDS}, we can assume without
loss of generality that for some distinct  $a$, $b$, $c$, and
$d$ from $\Sigma_{2^2}$ the MDS codes
 \begin{eqnarray*}
 T^a&=&\{(z_2,z_3,z_4) \mid (a,c,z_2,z_3,z_4,c,\ldots,c)\in M\}, \\
 T^b&=&\{(z_2,z_3,z_4) \mid (b,c,z_2,z_3,z_4,c,\ldots,c)\in M\}, \\
 T^d&=&\{(z_0,z_3,z_4) \mid (z_0,c,d,z_3,z_4,c,\ldots,c)\in M\}
 \end{eqnarray*}
are nonlinear and, moreover, $T^a \cup T^b$ is a linearization.
 Denote
 \begin{eqnarray*}
  M^a&=&\{(z_1,z_2,\ldots,z_n) \mid (a,z_1,z_2,\ldots,z_n)\in M\},\\
M^b&=&\{(z_1,z_2,\ldots,z_n) \mid (b,z_1,z_2,\ldots,z_n)\in M\},\\
M^d&=&\{(z_0,z_1,z_3,\ldots,z_n) \mid (z_0,z_1,d,z_3\ldots,z_n)\in M\},\mbox{ and } \\
M^c&=&\{(z_0,z_2,\ldots,z_n) \mid (z_0,c,z_2,\ldots,z_n)\in M\}.
 \end{eqnarray*}
By the hypothesis of the lemma and
Proposition~\ref{p:(semi)linear-from-2MDS}(1),
each of
$M^a$, $M^b$, $M^c$, $M^d$ is included in a unique linearization, say  $D^a$, $D^b$, $D^c$, $D^d$, respectively.
Let
 \begin{eqnarray*}
 \chi_{D^a}(z_1,\ldots,z_n)&=&\alpha_1(z_1)+\dots+\alpha_{n}(z_n)+\alpha,\\
 \chi_{D^b}(z_1,\ldots,z_n)&=&\alpha'_1(z_1)+\dots+\alpha'_{n}(z_n)+\alpha',\\
 \chi_{D^d}(z_0,z_1,z_3,\ldots,z_n)&=&\gamma_0(z_0)+\gamma_1(z_1)+\gamma_3(z_3)+\dots+\gamma_{n}(z_n)+\gamma,\\
 \chi_{D^c}(z_0,z_2,\ldots,z_n)&=&\delta_0(z_0)+\delta_2(z_2)+\dots+\delta_{n}(z_n)+\delta,
\end{eqnarray*}
where $\alpha$, $\alpha'$, $\gamma$, $\delta$ are constants
from $\{0,1\}$ and $\alpha_i$, $\alpha'_i$, $\gamma_i$, $\delta_i$
are the characteristic functions of cardinality-$2$ subsets of $\{[0,1],[1,0],[1,1]\}$ (such functions and
constants are defined uniquely for each linearization).

Consider the subcode $$M^{a,c}=\{(z_2,\ldots,z_n) \mid
(a,c,z_2,\ldots,z_n)\in M\}.$$ It is nonlinear, as the subcode $T^a$
is nonlinear, and semilinear by the hypothesis of the lemma. By
Proposition~\ref{p:(semi)linear-from-2MDS}(2), $M^{a,c}$ is a subset
of a unique linearization $D^{a,c}$. Since
$\chi_{D^{a}}(c,z_2,\ldots,z_n)=\chi_{D^{a,c}}(z_2,\ldots,z_n)=\chi_{D^{c}}(a,z_2,\ldots,z_n)$,
we see that $\alpha_i=\delta_i$ for $i\geq 2$. Similarly,
$\alpha'_i=\delta_i$ for $i\geq 2$.
%Hence,$\alpha'_i=\alpha_i$,$i\geq 2$.

Similarly, considering the nonlinear semilinear subcode
$$M^{c,d}=\{(z_0,z_3\ldots,z_n) \mid (z_0,c,d,z_3,\ldots,z_n)\in M\},$$
we establish $\delta_i=\gamma_i$ for $i\geq 3$.

Next, consider 
$$M^{a,d}=\{(z_1,z_3\ldots,z_n) \mid (a,z_1,d,z_3,\ldots,z_n)\in M\}.$$
We cannot state that it is not linear and is covered by a unique linearization. 
However, we state that the linearizations 
\begin{eqnarray*}
 D^{a,d} &=& \{(z_1,z_3,\ldots,z_n) \mid (z_1,d,z_3,\ldots,z_n)\in D^a\},
\quad\mbox{and}\\
D^{d,a} &=& \{(z_1,z_3,\ldots,z_n) \mid (a,z_1,z_3,\ldots,z_n)\in D^d\},
\end{eqnarray*}
which both include $M^{a,d}$, coincide.
Indeed, we have 
 \begin{eqnarray*}
 \chi_{D^{a,d}}(z_1,z_3,\ldots,z_n)&=&\alpha_1(z_1)+\alpha_2(d)+\alpha_3(z_3)+\dots+\alpha_{n}(z_n)+\alpha,\\
 \chi_{D^{d,a}}(z_1,z_3,\ldots,z_n)&=&\gamma_0(a)+\gamma_1(z_1)+\gamma_3(z_3)+\dots+\gamma_{n}(z_n)+\gamma,
\end{eqnarray*}
and we already know that $\alpha_3=\delta_3=\gamma_3$. 
By Proposition~\ref{p:two-double-codes}, 
this implies $D^{a,d}= D^{d,a}$ 
and, in particular $\alpha_1=\gamma_1$. 
Analogously, $\alpha'_1=\gamma_1$.
Together with previous results, we have $\alpha'_i=\alpha_i$ for $i\geq 1$,
which also means that $\alpha'=\alpha$ 
(indeed, 
$\chi_{D^a}(c,z_2,z_3,z_4,c,\ldots,c)
=\chi_{T^a\cup T^b}(z_2,z_3,z_4)
=\chi_{D^a}(c,z_2,z_3,z_4,c,\ldots,c)$).

Finally, we find that $M\subset D$, where
$$\chi_{D}(z_0,z_1,\dots,z_n)=\chi_{\{c,d\}}(z_0)+\alpha_1(z_1)+\dots+\alpha_{n}(z_n)+\alpha.$$
By Proposition~\ref{p:(semi)linear-from-2MDS}(1), the MDS code $M$ is semilinear.
\end{proof}

% By induction, it is easy to show the following.
% \begin{corollary}\label{c:semilinear-from-4-to-n}
% Any non-semilinear MDS code has a non-semilinear $4$-subcode.
% \end{corollary}

% \begin{corollary}
% An MDS code $M\subset \Sigma_{2^2}^n$ ($n\geq 5$) is semilinear if and
% only if all its $4$-subcodes are semilinear.
% \end{corollary}

\subsection{Every isotopically transitive latin hypercube of order $4$ is semilinear of degree at most $2$}
\begin{theorem}\label{th:order4}
A latin hypercube $g:\Sigma_{2^2}^n\to \Sigma_{2^2}$ is isotopically transitive if and
only if it is isotopic to a latin hypercube
(\ref{eq:1}) where $r$ is a Boolean function of degree at most $2$.
\end{theorem}
\begin{proof}
The ``if'' part is straightforward from Theorem~\ref{th:quadratic-topolinear}.
Let us show that there are no other isotopically transitive latin hypercubes of order $4$. 
Denote by $M$ the graph of $g$.

1) As follows by induction from Lemma~\ref{l:from-n-1-to-n}, 
every non-semilinear MDS code $M$
contains a non-semilinear $4$-subcode, which is not
isotopically transitive by Proposition~\ref{p:non-transitive-length4-MDS}.
 By Proposition~\ref{p:subcodes-of-IT}, 
 $M$ is not isotopically transitive either.

2) Assume that $g$ is a semilinear latin hypercube with a function $r$ of
degree at least $3$.
Considering the polynomial representation of $r$, we see that 
$r$ has a subfunction of degree $3$ in three
arguments.
The corresponding $4$-subcode of $M$ is isotopic 
to the semilinear MDS
code with the function $r_4$, see (\ref{eq:r4}).
As mentioned in Proposition~\ref{p:non-transitive-length4-MDS}, the last MDS
code is not isotopically transitive.
% Any such subfunction can be transformed
% to a function of form
% $r(x_1,x_2,x_3)=x_1x_2x_3+\delta_1x_1+\delta_2x_2+\delta_3x_3+\delta_0$
% by substituting $x_i+1$ instead of $x_i$ for some $i$. By addition
% of any linear function or a constant to $r$, we obtain a code
% isotopic to the initial code. Thus, any semilinear MDS code with
% function $r$ of degree at least $3$ has a subcode isotopic to $S_4$.
% As mentioned in Proposition~\ref{p:non-transitive-length4-MDS}, the semilinear MDS
% code $S_4$ with the function $r_4$ is not isotopically transitive.
By Proposition~\ref{p:subcodes-of-IT}, $M$ is not isotopically transitive either.
\end{proof}
\begin{corollary}\label{cor:order4}
Every isotopically transitive latin hypercube of order $4$ is topolinear.
\end{corollary}
%============================================================
\section{An isotopically transitive latin $n$-cube of order $5$ is unique}\label{s:5}
\begin{theorem}\label{th:5}
Any isotopically transitive latin $n$-cube of order $5$ is isotopic to the iterated group $Z_5$.  
\end{theorem}
\begin{proof}
  There are $15$ equivalence classes of latin $3$-cube of order $5$, see \cite{MK-W:small};
  representatives can be found in \cite{MK-W:data}, together with the orders of the autotopism groups. 
  One can see that only one latin $3$-cube, up to equivalence, has at least $5^3=125$ isotopisms, 
  and thus can be isotopically transitive.
  Of course, it is the iterated group.
  For an arbitrary $n$, the result follows from Theorem~8.8 in \cite{Zaslavsky:2012},
  which states (we give here a weakened version, in terms of MDS codes) 
  that if every $4$-subcode of an MDS code is a graph of an iterated group,
  then the MDS code is itself a graph of an iterated group.
\end{proof}

%============================================================
\section{Conclusion: open problems}\label{s:end}

It occurs that the isotopically transitive latin hypercubes considered in this paper
 are topolinear in most cases.
In particular, it follows from the characterization in Subsection~\ref{s:4}
that all isotopically transitive latin hypercubes of order $4$ are topolinear.
However, we did not establish the topolinearity of the latin hypercubes constructed in Theorem~\ref{th:iter1}
(the minimal example is the latin $3$-cube of order $6$ obtained as the composition of the loops $C_6$ and $D_6$).
In general, the following questions arise:
Does there exist a non-topolinear isotopically transitive latin hypercubes?
In particular, does there exist a non-topolinear isotopically transitive latin square?
The existence of nonpropelinear transitive codes with given parameters is not a simple problem in general; 
however, there are some positive results in this topic, for example, for perfect binary codes \cite{MogSol:2015:nonprope}.

Another natural problem is the characterization of 
isotopically transitive latin hypercubes of some given small order $q>5$,
say, $q=6$, $7$, $8$, $9$, which does not seem to be impossible.
It is also interesting to consider isotopically transitive latin hypercubes of all prime orders.
Hypothetically, they are equivalent to iterated cyclic groups. 
Note that any isotopically transitive latin hypercube of prime order has the following property: 
every $3$-subcode of its graph is equivalent to the graph of the cyclic group.
All latin hypercubes of orders $5$ and $7$ satisfying this property were considered in \cite{KroPot:retr},
where they are called sublinear. In particular Corollary~2 in that paper states 
that every sublinear latin $n$-cube of order $7$ is a composition of $n-1$ latin squares.

From the connection between the isotopically transitive latin squares and the G-loops established in Section~\ref{s:itG}
arises a question about the possibility to generalize this fact to more than two dimensions.
An $n$-ary loop is an algebraic system $(\Sigma,f)$, where $f$ is a latin $n$-cube satisfying 
$f(x,o,\ldots,o)=f(o,x,o,\ldots,o)=\dots=f(o,\ldots,o,x)=x$ for every $x$ and some $o$, called an identity element.
Similarly to the binary case, an $n$-ary loop $L$ can be called an $n$-ary G-loop if every $n$-ary loop isotopic
to $L$ is isomorphic to $L$. Then, the ``only if'' statement of Theorem~\ref{th:IT=G} will be still valid, 
with the same proof
(in particular, all constructions considered in the current paper give $n$-ary G-loops).
However, the proof of the ``if'' part does not work for $n\ge 3$. 
The reason is that an $n$-ary loop can have more than one 
identity element, and it is not clear at the moment 
if different identity elements necessarily generate the same orbit
under the autotopism group. 
The existence of $n$-ary G-loops that are not isotopically transitive remains an open question.

The last open problem we mention here is the construction 
of nonlinear transitive MDS codes of distance more than $2$.
Such objects are equivalent to systems of orthogonal latin hypercubes, 
with a specially defined orthogonality, see e.g. \cite{EthMul:2012}.
A partial case (length $4$) of this problem was considered in \cite{Pot:Maltsev15}.

\section{Acknowledgements}
The work was funded by the Russian Science Foundation (grant No 14-11-00555).

%\bibliographystyle{plain}
%\bibliography{../../k}

\begin{thebibliography}{10}

\bibitem{Albert:43}
A.~A. Albert.
\newblock Quasigroups. {I}.
\newblock {\em \href{http://www.ams.org/tran/}{Trans. Am. Math. Soc.}},
  54(3):507--519, 1943.
\newblock \DOI{10.1090/S0002-9947-1943-0009962-7}.

\bibitem{Andrews:partitions}
G.~E. Andrews.
\newblock {\em The Theory of Partitions}, volume~2 of {\em Encyclopedia of
  Mathematics and its Applications}.
\newblock Addison-Wesley, London, 1976.

\bibitem{Brouwer}
A.~E. Brouwer, A.~M. Cohen, and A.~Neumaier.
\newblock {\em Distance-Regular Graphs}.
\newblock Springer-Verlag, Berlin, 1989.
\newblock \DOI{10.1007/978-3-642-74341-2}.

\bibitem{Cher}
A.~V. Cheremushkin.
\newblock Canonical decomposition of $n$-ary quasigroups.
\newblock In {\em Issledovanie Operatsyj i Kvazigrupp}, volume 102 of {\em Mat.
  Issled.}, pages 97--105. Shtiintsa, Kishinev, 1988.
\newblock In Russian.

\bibitem{DonGra:2013}
D.~M. Donovan and M.~J. Grannell.
\newblock On the number of transversal designs.
\newblock {\em \href{http://www.sciencedirect.com/science/journal/00973165}{J.
  Comb. Theory, Ser.~A}}, 120(7):1562--1574, 2013.
\newblock \DOI{10.1016/j.jcta.2013.05.004}.

\bibitem{EthMul:2012}
J.~T. Ethier and G.~L. Mullen.
\newblock Strong forms of orthogonality for sets of hypercubes.
\newblock {\em
  \href{http://www.sciencedirect.com/science/journal/0012365X}{Discrete
  Math.}}, 312(12-13):2050--2061, 2012.
\newblock \DOI{10.1016/j.disc.2012.03.008}.

\bibitem{GR:82:Loops}
E.~G. Goodaire and D.~A. Robinson.
\newblock A class of loops which are isomorphic to all loop isotopes.
\newblock {\em \href{http://cms.math.ca/cjm/}{Can. J. Math.}}, 34(3):662--672,
  1982.
\newblock \DOI{10.4153/CJM-1982-043-2}.

\bibitem{Ito:patent}
T.~Ito.
\newblock Creation method of table, creation apparatus, creation program and
  program storage medium, 2004.
\newblock http://ip.com/patapp/US20040243621 New date stamp at
  http://www.freepatentsonline.com/7228311.html http://ip.com/patent/US7228311.

\bibitem{KKO:smallMDS}
J.~I. Kokkala, D.~S. Krotov, and P.~R.~J. {\"O}sterg{\aa}rd.
\newblock On the classification of {MDS} codes.
\newblock {\em
  \href{http://ieeexplore.ieee.org/xpl/RecentIssue.jsp?punumber=18}{IEEE Trans.
  Inf. Theory}}, 2015.
\newblock Accepted. \DOI{10.1109/TIT.2015.2488659}.

\bibitem{KokOst:Gr-Lat}
J.~I. Kokkala and P.~R.~J. {\"O}sterg{\aa}rd.
\newblock Classification of {G}raeco--{L}atin cubes.
\newblock {\em
  \href{http://onlinelibrary.wiley.com/journal/10.1002/%28ISSN%291520-6610}{J.
  Comb. Des.}}, 23(12):509--521, 2015.
\newblock \DOI{10.1002/jcd.21400}.

\bibitem{KokOst:further}
J.~I. Kokkala and P.~R.~J. {\"O}sterg{\aa}rd.
\newblock Further results on the classification of {MDS} codes.
\newblock E-print 1504.06982, arXiv.org, 2015.
\newblock Available at \url{http://arxiv.org/abs/1504.06982}.

\bibitem{KroPot:4}
D.~S. Krotov and V.~N. Potapov.
\newblock $n$-{A}ry quasigroups of order $4$.
\newblock {\em \href{http://epubs.siam.org/journal/sjdmec}{SIAM J. Discrete
  Math.}}, 23(2):561--570, 2009.
\newblock \DOI{10.1137/070697331}.

\bibitem{KroPot:retr}
D.~S. Krotov and V.~N. Potapov.
\newblock On connection between reducibility of an $n$-ary quasigroup and that
  of its retracts.
\newblock {\em
  \href{http://www.sciencedirect.com/science/journal/0012365X}{Discrete
  Math.}}, 311(1):58--66, 2011.
\newblock \DOI{10.1016/j.disc.2010.09.023}.

\bibitem{KroPot:2012:propel}
D.~S. Krotov and V.~N. Potapov.
\newblock Propelinear $1$-perfect codes from quadratic functions.
\newblock {\em
  \href{http://ieeexplore.ieee.org/xpl/RecentIssue.jsp?punumber=18}{IEEE Trans.
  Inf. Theory}}, 60(4):2065--2068, 2014.
\newblock \DOI{10.1109/TIT.2014.2303158}.

\bibitem{KPS:ir}
D.~S. Krotov, V.~N. Potapov, and P.~V. Sokolova.
\newblock On reconstructing reducible $n$-ary quasigroups and switching
  subquasigroups.
\newblock {\em \href{http://www.math.md/en/publications/qrs/}{Quasigroups
  Relat. Syst.}}, 16(1):55--67, 2008.
\newblock Online:
  \url{http://www.quasigroups.eu/contents/download/2008/16_7.pdf}.

\bibitem{Kunen:99}
K.~Kunen.
\newblock G-loops and permutation groups.
\newblock {\em \href{http://www.sciencedirect.com/science/journal/00218693}{J.
  Algebra}}, 220(2):694--708, 1999.
\newblock \DOI{10.1006/jabr.1999.8006}.

\bibitem{LinLur:2014}
N.~Linial and Z.~Luria.
\newblock An upper bound on the number of high-dimensional permutations.
\newblock {\em \href{http://link.springer.com/journal/493}{Combinatorica}},
  2014.
\newblock \DOI{10.1007/s00493-014-2842-8}.

\bibitem{MK-W:data}
B.~D. McKay and I.~M. Wanless.
\newblock Combinatorial data. {L}atin cubes and hypercubes.
\newblock
  \url{https://cs.anu.edu.au/people/Brendan.McKay/data/latincubes.html}.

\bibitem{MK-W:small}
B.~D. McKay and I.~M. Wanless.
\newblock A census of small {L}atin hypercubes.
\newblock {\em \href{http://epubs.siam.org/journal/sjdmec}{SIAM J. Discrete
  Math.}}, 22(2):719--736, 2008.
\newblock \DOI{10.1137/070693874}.

\bibitem{MogSol:2015:nonprope}
I.~{Yu}. Mogilnykh and F.~I. Solov'eva.
\newblock Transitive nonpropelinear perfect codes.
\newblock {\em
  \href{http://www.sciencedirect.com/science/journal/0012365X}{Discrete
  Math.}}, 338(3):174--182, 2015.
\newblock \DOI{10.1016/j.disc.2014.11.001}.

\bibitem{Pot:trans}
V.~N. Potapov.
\newblock A lower bound for the number of transitive perfect codes.
\newblock {\em
  \href{http://link.springer.com/journal/volumesAndIssues/11754}{J. Appl. Ind.
  Math.}}, 1(3):373--379, 2007.
\newblock \DOI{10.1134/S199047890703012X} translated from
  \href{http://www.mathnet.ru/php/journal.phtml?jrnid=da\&option_lang=eng}{Diskretn.
  Anal. Issled. Oper.}, Ser.~1, 13(4):49-59, 2006.

\bibitem{Pot:2012:partial}
V.~N. Potapov.
\newblock On extensions of partial $n$-quasigroups of order $4$.
\newblock {\em \href{http://link.springer.com/journal/12002}{Sib. Adv. Math.}},
  22(2):135--151, 2012.
\newblock \DOI{10.3103/S1055134412020058} translated from
  \href{http://www.mathnet.ru/php/journal.phtml?jrnid=mt\&option_lang=eng}{Mat.
  Tr.} 14(2):147-172, 2012.

\bibitem{Pot:Maltsev15}
V.~N. Potapov.
\newblock Isotopically transitive pairs of {MOLS}.
\newblock In {\em International Conference ``{M}al'tsev Meeting''. May 3--7,
  2015. Collection of Abstracts}, page 146, Novosibirsk, 2015. Sobolev
  Institute of Mathematics, Novosibirsk State University.
\newblock Online: \url{http://math.nsc.ru/conference/malmeet/15/malmeet15.pdf}.

\bibitem{Pot:Number-MDS}
V.~N. Potapov.
\newblock On the number of latin hypercubes, pairs of orthogonal latin squares
  and {MDS} codes.
\newblock E-print 1510.06212, arXiv.org, 2015.
\newblock Available at \url{http://arxiv.org/abs/1510.06212}.

\bibitem{PotKro:numberQua.en}
V.~N. Potapov and D.~S. Krotov.
\newblock On the number of $n$-ary quasigroups of finite order.
\newblock {\em \href{http://www.degruyter.de/journals/dma/detail.cfm}{Discrete
  Math. Appl.}}, 21(5-6):575--585, 2011.
\newblock DOI{10.1515/dma.2011.035}, translated from
  \href{http://www.sciencedirect.com/science/journal/0012365X}{Discrete Math.}
  24:1 (2012), 60--69.

\bibitem{RBH:89}
J.~Rif\`a, J.~M. Basart, and L.~Huguet.
\newblock On completely regular propelinear codes.
\newblock In {\em Applied Algebra, Algebraic Algorithms and Error-Correcting
  Codes, 6th Int. Conference, AAECC-6 Rome, Italy, July 4--8, 1988
  Proceedings}, volume 357 of {\em
  \href{http://link.springer.com/bookseries/558}{Lect. Notes Comput. Sci.}},
  pages 341--355. Springer-Verlag, 1989.
\newblock \DOIURL{10.1007/3-540-51083-4\_71}{10.1007/3-540-51083-4_71}.

\bibitem{Wilson:74}
R.~L. Wilson, Jr.
\newblock Isotopy-isomorphy loops of prime order.
\newblock {\em \href{http://www.sciencedirect.com/science/journal/00218693}{J.
  Algebra}}, 31(1):117--119, 1974.
\newblock \DOI{10.1016/0021-8693(74)90008-8}.

\bibitem{Wilson:75}
R.~L. Wilson, Jr.
\newblock Quasidirect products of quasigroups.
\newblock {\em \href{http://www.tandfonline.com/loi/lagb20}{Commun. Algebra}},
  3(9):835--850, 1975.
\newblock \DOI{10.1080/00927877508822075}.

\bibitem{Zaslavsky:2012}
T.~Zaslavsky.
\newblock Associativity in multary quasigroups: the way of biased expansions.
\newblock {\em \href{http://link.springer.com/journal/10}{Aequationes Math.}},
  83(1-2):1--66, 2012.
\newblock \DOI{10.1007/s00010-011-0086-x}.

\bibitem{ZZ:2002}
V.~A. Zinoviev and D.~V. Zinoviev.
\newblock Binary extended perfect codes of length 16 by the generalized
  concatenated construction.
\newblock {\em \href{http://link.springer.com/journal/11122}{Probl. Inf.
  Transm.}}, 38(4):296--322, 2002.
\newblock \DOI{10.1023/A:1022049912988} translated from
  \href{http://www.mathnet.ru/php/journal.phtml?jrnid=ppi\&option_lang=eng}{Probl.
  Peredachi Inf.} 38(4) (2002), 56-84.

\end{thebibliography}
%\end{document}

\providecommand\href[2]{#2} \providecommand\url[1]{\href{#1}{#1}}
  \providecommand\bblmay{May} \providecommand\bbloct{October}
  \providecommand\bblsep{September} \def\DOI#1{{\small {DOI}:
  \href{http://dx.doi.org/#1}{#1}}}\def\DOIURL#1#2{{\small{DOI}:
  \href{http://dx.doi.org/#2}{#1}}}\providecommand\bbljun{June}

\end{document}